# Spread of Information with Confirmation Bias in Cyber-Social Networks

Yanbing Mao, Sadegh Bolouki, and Emrah Akyol

**Abstract**—This paper provides a model to investigate information spreading over cyber-social network of agents communicating with each other. The cyber-social network considered here comprises of individuals and news agencies. Each individual holds a belief represented by a scalar. Individuals receive information from news agencies that are closer to their belief, confirmation bias is explicitly incorporated into the model. The proposed dynamics of cyber-social networks is adopted from DeGroot-Friedkin model, where the individual's opinion update mechanism is a convex combination of his innate opinion, his neighbors' opinions at the previous time step (obtained from the social network), and the opinions passed along by news agencies from cyber layer which he follows. The characteristics of the interdependent social and cyber networks are radically different here: the social network relies on trust and hence static while the news agencies are highly dynamic since they are weighted as a function of the distance between an individual state and the state of news agency to account for confirmation bias. The conditions for convergence of the aforementioned dynamics to a unique equilibrium are characterized. The estimation and exact computation of the steady-state values under non-linear and linear state-dependent weight functions are provided. Finally, the impact of polarization in the opinions of news agencies on the public opinion evolution is numerically analyzed in the context of the well-known Krackhardt's advice network.

## 1 INTRODUCTION

INDIVIDUALS form belief on various social, political and economic issues based on three factors: innate opinion that is based on inherent personal characteristics (e.g., socio-economic conditions in which the individual grew up and/or live in); the information they receive from other individuals, including friends, coworkers; the information received from the news sources and followed thought leaders, see e.g., [1]–[4]. Typically, a social network (friends, neighbors, coworkers) is informationally *symmetric* and *static*: when people exchange information, they get influence, and also get influenced by others; and the connectivity is based on a static (information sense) network that does not depend on the individual beliefs[1]. However, the interaction between the public and news sources (or thought leaders) is informationally asymmetric, in the sense that news agencies will not update their beliefs by the information they receive from public (information exchange is only one directional: from news agencies to the public). Moreover, the connectivity of the communication network between a news source and an individual depends on the current state of the aforementioned individual and the news agency. This well-known fact, namely the **confirmation bias** [5] prevalent in the current societies, makes this network highly dynamic: the set of news agencies that an individual gets information from depend on the distance between the values of states of the individual and the news agency. This entire system is an example of a cyber-social network comprised of two interacting, interdependent networks with radically different characteristics: an informationally symmetric, static social network and an asymmetric, dynamic cyber network.

In this paper, we study the dynamics of information spread on such cyber-social networks, with a particular focus on confirmation bias. Confirmation bias refers to a type of cognitive bias that involves favoring information which confirms previously existing beliefs or biases. It is well understood that machine learning algorithms that control the information on social media news feeds automatically utilize and foster this bias without the individual's permission or even proper understanding, see e.g., [6] and the recent news articles at popular media outlets [7], [8].

We note that there exist substantial amount of prior work on network dynamics, we summarize a few popular models in Table 1. For example, DeGroot model [1] is inspired by the idea of relieving psychological discomfort from their disagreement with others; DeGroot-Friedkin model [2] includes the evolution of self-confidence levels after discussion on issues; Hegselmann-Krause model [3] consider a complex situation where an agent will take into account the opinions of others to a certain extent in forming his own opinion.

The recent research focus has shifted to variations of such classical models in order to capture the subtle characteristics of actual social networks. For example, the studies [9], [10] allow the individuals to have self-appraisal mechanism through updating of individuals' self-confidence level after discussion of issues. Dhamal et al. [11] incorporate opponent stubborn agents into DeGroot-Friedkin model to study competitive information spreading in social networks. DeGroot-Friedkin model has been used in several problems in social networks, including de-biasing social wisdom in online social networks [12], the competitive propagation in social networks [11], the optimal opinion conformation [13], and many more. A variant of the DeGroot-Friedkin model also appears as a game-theoretic best response dynamics for a specific potential game [14]. There has been a renewed interest also in modeling the manipulation behavior and misinformation (or fake news) spread over the networks via network and/or game theoretic tools, see e.g., [11], [15]–[19]. We note that our model can also represent the fake news spread, for the special case that an adversary controls a subset of the news agencies. However, the information dynamics models in the prior work do not adequately capture the characteristics of the current and emerging cyber-social networks that typically involves multiple interdependent, interacting networks with different properties. The following features differentiates our model from the prior work:

1) we particularly focus on the steady-state (equilibrium) where agents converge to different state values; while most of the prior work study consensus, that is all agents converge to the same state. We believe it might be hard to achieve consensus in the presence of polarized opinions and confirmation bias; and hence our model captures reality more accurately.
2) individual's opinion update mechanism is a convex combination of her innate opinion, the opinions of her neighbors at the previous time step (social network), and the opinions carried by news agencies that she follows (influence from the cyber layer);
3) the weight of influence of individuals is fixed because social influence among individuals is based on "trust", which tends to vary little over a long period of time, while the weight of influence of news agencies over individuals heavily depend on the current opinions of individuals, i.e., it is state-dependent, in order to capture the impact of "confirmation bias".

---

• Y. Mao, S. Bolouki and E. Akyol are with the Department of Electrical and Computer Engineering, Binghamton University–SUNY, Binghamton, NY, 13902 USA. E-mail: {ymao3, eakyol}@binghamton.edu; sadegh.bolouki@gmail.com. A slightly condensed version of this paper is submitted to the 57th IEEE Conference on Decision and Control, 2018.

1. In this paper, we use "belief", "opinion" and "state" interchangeably.

The contribution of this paper is threefold, which can be summarized as follows.

- Based on the well-known DeGroot-Friedkin model, we propose a dynamics of cyber-social networks, where the weight of influence of individuals is fixed, which is inspired by the little varying "trust" in friendship network, while the weight of influence of news agencies is state-dependent (to capture confirmation bias), which is inspired by the idea of Hegselmann-Krause model. The conditions for the convergence to a unique equilibrium point are characterized.
- The estimation and the exact computation (if possible) of equilibrium point of the proposed dynamics under linear and nonlinear state-dependent weight functions are provided.
- Using the proposed dynamics of cyber-social networks, the effects of the distribution of news agency's opinions and the distance between polarized opinions of news agencies are studied in the context of the well-known Krackhardt's advice network [20]. The numerical results demonstrate that 1) influencing one critical individual by news agency can result in largest sample deviation; 2) the order of influences of individuals that follow news agency on sample de-viation are preserved under different uniform distributions; 3) bi-model uniform distribution of opinion of new agency yields larger sample deviation than a single-mode uniform distribution; 4) the longer distance between the means of polar opinions results in the bigger sample variance.

Table 1
Social Network Models

| Ref. | Dynamics | Name |
|---|---|---|
| [1] | $x(t+1) = Wx(t)$ | DeGroot model |
| [2] | $x(t+1) = Ax(t) + (\mathbf{1} - A)s$ | DeGroot-Friedkin model |
| [3] | $\begin{cases} x(t+1) = A(x(t))x(t), \\ w_{ij}(x) = \begin{cases} > 0, |x_i - x_j| < \varepsilon \\ = 0, |x_i - x_j| \geq \varepsilon \end{cases} \end{cases}$ | Hegselmann-Krause model |

This paper is organized as follows. In Section 2, we present the preliminaries and problem formulation. In Sections 3 and 4, we analyze the convergence of dynamics and the equilibrium (steady-state) point respectively. We provide numerical results in Section 5, and in Section 6 we present conclusions and future research directions.

## 2 PRELIMINARIES AND PROBLEM FORMULATION
### 2.1 Notation and Terminology

We let $\mathbb{R}^n$ and $\mathbb{R}^{m \times n}$ denote the set of $n$-dimensional real vectors and the set of $m \times n$-dimensional real matrices, respectively. Given a vector $x \in \mathbb{R}^n$ and a matrix $A \in \mathbb{R}^{n \times m}$, inequalities $x \geq 0$ and $A \geq 0$ denote element-wise inequalities. $\mathbb{N}$ represents the set of the natural numbers and $\mathbb{N}_0 = \mathbb{N} \cup \{0\}$. We let $\mathbf{1}$ and $\mathbf{0}$ be the identity and zero matrices with proper dimensions, respectively. We let $\mathbf{1}_n \in \mathbb{R}^n$ and $\mathbf{0}_n \in \mathbb{R}^n$ denote the vector of all ones and all zeros, respectively. The superscript '$\top$' stands for the matrix transposition. A square matrix with non-negative entries is said to be sub-stochastic (or strictly sub-stochastic) if the entries of each row of the matrix sum up to one or less (or less than one).

The network considered in this paper is composed of $n$ individuals and $m$ news agencies. The interaction among the individuals is modeled by a digraph $\mathcal{G} = (\mathcal{V}, \mathcal{E})$, where $\mathcal{V} = \{v_1, \cdots, v_n\}$ is the set of vertices representing the individuals and $\mathcal{E} \subset \mathcal{V} \times \mathcal{V}$ is the set of edges of the digraph $\mathcal{G}$ representing the influence structure. The communication from news agencies to individuals is modeled by a bipartite digraph $\mathcal{H} = (\mathcal{V} \bigcup \mathcal{K}, \mathcal{B})$, where $\mathcal{K} = \{u_1, \cdots, u_m\}$ is the set of vertices representing the news agencies and $\mathcal{B} \subset \mathcal{V} \times \mathcal{K}$ is the set of edges of the digraph. The adjacency matrix $B = [b_{ik}] \in \mathbb{R}^{n \times m}$ of the digraph $\mathcal{H}$ is defined as $b_{ik} = 1$ if new agency $u_k$ has influence on individual $v_i$, and $b_{ik} = 0$ otherwise.

Some important notations are highlighted as follows:

| | |
|---|---|
| $\mathcal{V}$ | the set of individuals; |
| $\mathcal{K}$ | the set of news agencies; |
| $\|\cdot\|$ | $l_1$ norm of a vector, and the induced norm of matrix $A \in \mathbb{R}^{n \times n}$, i.e., $\|A\| = \sup\{\frac{\|Ax\|}{\|x\|} : x \in \mathbb{R}^n \text{ with } x \neq \mathbf{0}_n\}$; |
| $[x]_i$ | the $i^{\text{th}}$ entry of vector $x \in \mathbb{R}^n$ (the opinion of the $i^{\text{th}}$ agent); |

$\mathcal{W}(d)$    the $d^{\text{th}}$ element in the given ordered set $\mathcal{W}$;
$E_{\text{U}}(\cdot)$    the expectation operator over distribution U.

## 2.2 Auxiliary Lemma

The following well-known result will be used throughout the paper to prove the convergence of dynamics to a unique equilibrium.

***Lemma 1 (Banach fixed-point theorem [21]).*** Let $(X;d)$ be a complete metric space and $f: X \to X$ be a map such that $d(f(x); f(x')) \leq c d(x; x')$ for some $0 < c < 1$ and all $x$ and $x'$ in $X$. Then $f$ has a unique fixed point in $X$. Moreover, for any $x_0 \in X$ the sequence of iterates $x_0$; $f(x_0); f(f(x_0)); \ldots$ converges to the fixed point of $f$.

## 2.3 Problem Formulation

For convenience, we refer to extremely stubborn agents that do not change their opinion as "news agencies" (cyber-layer). Other agents are simply referred to as "individuals" and update their opinion based on their neighbors and news agencies. We consider the following model which is adopted from the DeGroot-Friedkin model [2]:

$$x_i(t+1) = \alpha_i(x_i(t)) s_i + \sum_{j \in \mathcal{V}} w_{ij} x_j(t) + \sum_{k \in \mathcal{K}} \hat{w}_{ik}(x_i(t)) y_k, \quad (1)$$

where

I) $x_i(t) \in [0,1]$ is individual $v_i$'s opinion at time $t$, $s_i$ is his fixed innate opinion, while $y_k$ is the news agency $u_k$'s opinion;

II) $\alpha_i(x_i)$, is referred to as the "resistance parameter" of individual $v_i$ is determined in such a way that it satisfies

$$\alpha_i(x_i(t)) + \sum_{j \in \mathcal{V}} w_{ij} + \sum_{k \in \mathcal{K}} \hat{w}_{ik}(x_i(t)) = 1, \forall i \in \mathcal{V}, \forall t. \quad (2)$$

III) $w_{ij}$ represents the weighted influence of individual $v_j$ on individual $v_i$,

$$\begin{cases} w_{ij} > 0 & \text{if } (v_i, v_j) \in \mathcal{E} \\ w_{ij} = 0 & \text{otherwise.} \end{cases} \quad (3)$$

We note that $w_{ij}$ does not depend on time index $t$.

IV) $\hat{w}_{ik}(x_i(t))$ is the weighted influence of news agency $u_k$ on individual $v_i$ with

$$\hat{w}_{ik}(x_i(t)) = \begin{cases} g_{ik}(|x_i(t) - y_k|), & b_{ik} = 1 \\ 0, & b_{ik} = 0 \end{cases} \quad (4)$$

where $g_{ik}(|x_i(t) - y_k|) : \mathbb{R} \to \mathbb{R}$, is a strictly decreasing function with respect to $|x_i(t) - y_k|$, and it satisfies $1 > g_{ik}(|x_i(t) - y_k|) > 0$ for $\forall i \in \mathcal{V}, \forall k \in \mathcal{K}, \forall t \in \mathbb{N}_0$.

We next make the following assumption on the weight functions in (4).

***Assumption 1.*** The weight function $g_{ik}(\cdot) : \mathbb{R} \to \mathbb{R}$ in (4) satisfies

$$|g_{ik}(z_1) - g_{ik}(z_2)| \leq \mu_i |z_1 - z_2|, \quad (5)$$

for some fixed $\mu_i \in \mathbb{R}$.

***Remark 1 (Motivation of Weight of Influence).*** We assume that $w_{ij}$'s are fixed because social influence among individuals is based on "trust", which tends to vary little over a long period of time. However, the influence of news agencies over individuals depend heavily on the current opinions of individuals, due to the confirmation bias. For instance, Facebook and Twitter famously incorporate the confirmation bias in placing the news in their individualized newsfeed. That is why the weight of influence of news agency on individual, i.e., $\hat{w}_{ik}(x_i(t))$ defined as (4) is state-dependent.

***Remark 2.*** The sum of coefficients equal to one, i.e, the condition (2), is a standard practice in modeling opinion evolution (see e.g., [11]), since the dynamics are invariant under translation. Here, (2) shows the evolution of an individual $v_i$'s opinion at each time step is a convex combination of his innate opinion $s_i$, his neighbors' opinions at the previous time step, and the opinions $y_k$ passed along by news agencies which he follows. Noting that the innate opinions, the initial opinions $(x_i(0), i \in \mathcal{V})$, and the opinions of news agencies all belong to the $[0,1]$, by iteration from (1) we have $x_i(t) \in [0,1]$ for $\forall t \in \mathbb{N}_0, \forall i \in \mathcal{V}$.

***Remark 3.*** With the fact $x_i(t) \in [0,1]$ for $\forall t \in \mathbb{N}_0, \forall i \in \mathcal{V}$ stated in Remark 2, the condition (5) in Assumption 1 is not a restrictive assumption condition on weight functions. It allows the weight functions to be nonlinear. The following two examples illustrate this point.

***Example 1.*** Consider the function $g_{ik}(\cdot) : \mathbb{R} \to \mathbb{R}$,

$$g_{ik}\left(|\tilde{\tilde{x}}_i - y_k|\right) = \mu_i \ln\left(2 - |y_k - \tilde{\tilde{x}}_i|\right). \quad (6)$$

Without loss of generality, let $|y_k - \tilde{\tilde{x}}_i| \leq |y_k - z_i|$. It follows from the inequality $z \geq \ln(1+z)$ with $z > 0$, the facts $|y_k - z_i| \leq 1$ and $|y_k - \tilde{\tilde{x}}_i| \leq 1$ that

$$|g_{ik}(|\tilde{\tilde{x}}_i - y_k|) - g_{ik}(|z_i - y_k|)|$$
$$= \mu_i \log\left(1 + \frac{|y_k - z_i| - |y_k - \tilde{\tilde{x}}_i|}{2 - |y_k - z_i|}\right)$$
$$\leq \mu_i \frac{|y_k - z_i| - |y_k - \tilde{\tilde{x}}_i|}{2 - |y_k - z_i|}$$
$$\leq \mu_i \frac{|(y_k - z_i) - (y_k + \tilde{\tilde{x}}_i)|}{1 + 1 - |y_k - z_i|} \leq \mu_i |\tilde{\tilde{x}}_i - z_i|. \quad (7)$$

***Example 2.*** Consider the function $g_{ik}(\cdot) : \mathbb{R} \to \mathbb{R}$,

$$g_{ik}\left(|\tilde{\tilde{x}}_i - y_k|\right) = \beta_i - \mu_i \sin\left(|y_k - \tilde{\tilde{x}}_i|\right). \quad (8)$$

Considering $z \geq \sin(z)$ with $z \geq 0$, and the well known trigonometric identity $\sin \varpi - \sin \tau = 2 \cos \frac{\varpi + \tau}{2} \sin \frac{\varpi - \tau}{2}$, we have

$$|g_{ik}(|\tilde{\tilde{x}}_i - y_k|) - g_{ik}(|z_i - y_k|)|$$
$$= |\mu_i(\sin|y_k - \tilde{\tilde{x}}_i| - \sin|y_k - z_i|)|$$
$$= 2\mu_i |\cos \frac{\varpi + \tau}{2} \sin \frac{\varpi - \tau}{2}|$$
$$\leq 2\mu_i |\sin \frac{|y_k - \tilde{\tilde{x}}_i| - |y_k - z_i|}{2}|$$
$$\leq \mu_i |(|y_k - \tilde{\tilde{x}}_i|) - (|y_k - z_i|)| \leq \mu_i |\tilde{\tilde{x}}_i - z_i|, \quad (9)$$

where $\varpi = |y_k - \tilde{\tilde{x}}_i|$ and $\tau = |y_k - z_i|$.

## 3 CONVERGENCE ANALYSIS

This section studies the equilibrium point and the convergence of the social dynamics (1). Let us rewrite the dynamics (1) in the following matrix form:

$$x(t+1) = \alpha(x(t))s + Wx(t) + \widetilde{W}(x(t))y, \quad (10)$$

where we define the following variables:

$$s = [s_1, s_2, \cdots, s_n]^\top \in \mathbb{R}^n, \quad (11)$$
$$x(t) = [x_1(t), x_2(t), \cdots, x_n(t)]^\top \in \mathbb{R}^n, \quad (12)$$
$$y = [y_1, \cdots, y_m]^\top \in \mathbb{R}^m, \quad (13)$$
$$\alpha(x(t)) = \mathrm{diag}\{\alpha_1(x_1(t)), \cdots, \alpha_n(x_n(t))\} \in \mathbb{R}^{n \times n}, \quad (14)$$

$$W = \begin{bmatrix} w_{11} & \cdots & w_{1n} \\ w_{21} & \cdots & w_{2n} \\ \vdots & \vdots & \vdots \\ w_{n1} & \cdots & w_{nn} \end{bmatrix} \in \mathbb{R}^{n \times n}, \quad (15)$$

$$\widetilde{W}(x(t)) = \begin{bmatrix} \hat{w}_{11}(x_1(t)) & \cdots & \hat{w}_{1m}(x_1(t)) \\ \vdots & \vdots & \vdots \\ \hat{w}_{n1}(x_n(t)) & \cdots & \hat{w}_{nm}(x_n(t)) \end{bmatrix} \in \mathbb{R}^{n \times m}, \quad (16)$$

with $\hat{w}_{ik}(x_i(t)), i \in \mathcal{V}, k \in \mathcal{K}$ given in (4). In following theorem, we present the conditions for which the dynamics converge to a unique steady-state equilibrium.

***Theorem 1.*** Under Assumption 1, the dynamics converges to a unique equilibrium point if

$$\|W\| + 2\gamma < 1, \quad (17)$$

where

$$\gamma = \max_{i=1,\cdots,n}\{\mu_i \Gamma_i\}, \quad (18a)$$
$$\Gamma_i = \sum_{k \in \mathcal{K}} b_{ik}, \quad (18b)$$

*Proof:* Let us consider the matrix form (10) of the dynamics (1). Choose two vectors $\tilde{\hat{x}} \in \mathbb{R}^n$ and $z \in \mathbb{R}^n$. Let us first define

$$f(z) \triangleq \alpha(z)s + Wz + \widetilde{W}(z)y.$$

Then, it follows from (10) that

$$f(\tilde{\hat{x}}) - f(z) = \left(\alpha(\tilde{\hat{x}}) - \alpha(z)\right)s + W(\tilde{\hat{x}} - z) + \left(\widetilde{W}(\tilde{\hat{x}}) - \widetilde{W}(z)\right)y. \quad (19)$$

Recalling the definition of induced norm of a matrix $W \in \mathbb{R}^{n \times n}$ in Section 2.1, we have

$$\left\|W(\tilde{\hat{x}} - z)\right\| \leq \|W\|\|\tilde{\hat{x}} - z\|. \quad (20)$$

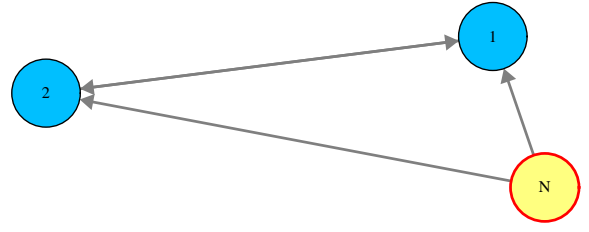

Figure 1. Communication Topology.

Under Assumption 1, and recalling that $1 \geq y_k \geq 0$, $b_{ik} = 1$ or 0, and $\mu_i \geq 0$, we obtain from (13), (16) and (4) that

$$\left|\left[(\widetilde{W}(\tilde{\hat{x}}) - \widetilde{W}(z))y\right]_i\right|$$
$$= \left|\sum_{k \in \mathcal{K}} b_{ik}(g_{ik}(|\tilde{\hat{x}}_i - y_k|) - g_{ik}(|z_i - y_k|))y_k\right|$$
$$\leq \left|\sum_{k \in \mathcal{K}} b_{ik}(g_{ik}(|\tilde{\hat{x}}_i - y_k|) - g_{ik}(|z_i - y_k|))y_k\right|$$
$$\leq \left|\tilde{\hat{x}}_i - z_i\right|\mu_i \sum_{k \in \mathcal{K}} b_{ik} y_k$$
$$\leq \left|\tilde{\hat{x}}_i - z_i\right|\mu_i \sum_{k \in \mathcal{K}} b_{ik}$$
$$= \left|\tilde{\hat{x}}_i - z_i\right|\mu_i \Gamma_i, i \in \mathcal{V}$$

which implies

$$\left\|\left(\widetilde{W}(\tilde{\hat{x}}) - \widetilde{W}(z)\right)y\right\| \leq \gamma \left\|\tilde{\hat{x}} - z\right\|, \quad (21)$$

where $\gamma$ and $\Gamma_i$ are given in (18). We note that (2) and (4) imply

$$\alpha_i(\tilde{\hat{x}}_i) - \alpha_i(z_i) = \sum_{k \in \mathcal{K}} \hat{w}_{ik}(z_i) - \sum_{k \in \mathcal{K}} \hat{w}_{ik}(\tilde{\hat{x}}_i).$$

Under Assumption 1, it follows from (11), (14) and the fact $0 \leq s_i \leq 1$ that

$$\left|\left[(\alpha(\tilde{\hat{x}}) - \alpha(z))s\right]_i\right|$$
$$= \left|\sum_{k \in \mathcal{K}} s_i b_{ik}(g_{ik}(|\tilde{\hat{x}}_i - y_k|) - g_{ik}(|z_i - y_k|))\right|$$
$$\leq \left|\tilde{\hat{x}}_i - z_i\right| s_i \mu_i \sum_{k \in \mathcal{K}} b_{ik}$$
$$\leq \left|\tilde{\hat{x}}_i - z_i\right|\mu_i \Gamma_i, \forall i \in \mathcal{V}$$

which implies

$$\left\|\left(\alpha(\tilde{\hat{x}}) - \alpha(z)\right)s\right\| \leq \gamma \left\|\tilde{\hat{x}} - z\right\|. \quad (22)$$

Combining (19) with (20), (21) and (22), we have

$$\left\|f(\tilde{\hat{x}}) - f(z)\right\| \leq (\|W\| + 2\gamma)\left\|\tilde{\hat{x}} - z\right\|.$$

If (17) holds, the condition in Lemma 1 would be satisfied, hence by Lemma 1, the dynamics (10) converges to a unique equilibrium point for any initial opinion $x(0) \in \mathbb{R}^n$, which completes the proof. □

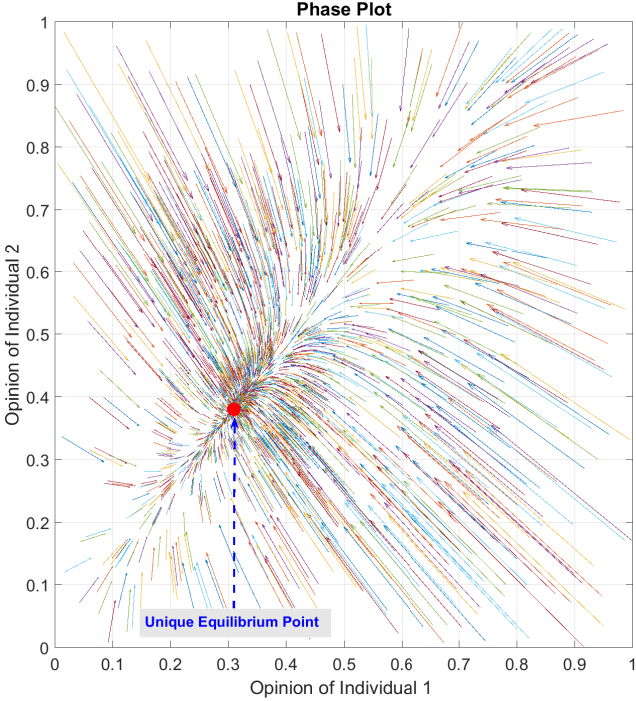

Figure 2. Evolution of individuals' opinions under 500 random initial opinions.

We next note the following important observation from Theorem 1: the asymptotic behavior of the system is independent of initial opinion $x(0) \in \mathbb{R}^n$, as illustrated by the example below.

*Example 3.* Consider a network of two individuals and one news agency, where its communication structure is give in Figure 1 with N indicting the news agency. Its detailed dynamics is described by

$$x_1(t+1) = (1 - 0.4 - 0.3(1 - |x_1(t) - y_1|))s_1 \quad (23a)$$
$$+ 0.4x_2(t) + 0.3(1 - |x_1(t) - y_1|)y_1,$$
$$x_2(t+1) = (1 - 0.6 - 0.4(1 - |x_2(t) - y_1|))s_2 \quad (23b)$$
$$+ 0.6x_1(t) + 0.4(1 - |x_2(t) - y_1|)y_1.$$

Set the individuals innate opinions as $[s_1, s_2] = [0.1, 0.4]$. The opinion forwarded by news agency is chosen to be $y_1 = 0.5$. The linear state-dependent weight functions satisfy condition (5) in Assumption 1. Moreover, the weight parameters of the social network satisfy (17). The phase plot of the two individuals' opinions under 500 random initial opinions is shown in Figure 2, which shows that the opinions of individuals converge to a unique equilibrium point, which is independent of initial opinion.

## 4 EQUILIBRIUM POINT

This section studies the impact of news agencies on the evolution of individuals' opinions.

### 4.1 Equilibrium Point Expression

We use $x^{\text{ue}}$ and $x^{\text{ie}}$ to denote the uninfluential equilibrium point, i.e., the equilibrium point of the system in the absence of news agencies, and influential equilibrium point, i.e., the equilibrium point of the system in the presence of news agencies, respectively.

#### 4.1.1 In the Absence of News Agencies

From (10), the dynamics of the system in the absence of news agencies is expressed as

$$\hat{x}(t+1) = \hat{\alpha}s + W\hat{x}(t), \quad (24)$$

where $W$ is given in (15) and

$$\hat{\alpha} = \text{diag}\{1 - \sum_{j \in \mathcal{V}} w_{1j}, \cdots, 1 - \sum_{j \in \mathcal{V}} w_{nj}\}. \quad (25)$$

*Corollary 1.* The social dynamics (24) converges to a unique equilibrium point

$$x^{\text{ue}} = (\mathbf{1} - W)^{-1} \hat{\alpha} s, \quad (26)$$

where $\hat{\alpha}$ is defined in (25).

*Remark 4.* The authors in [12] obtains nearly the same solution as (26). However, the weight matrix $W$ considered therein is a *strictly* sub-stochastic matrix while $W$ defined in (15) is a sub-stochastic matrix, the dynamics (24) still converges to a unique equilibrium point. Its brief proof is sketched as follows.

*Proof of Corollary 1:* Note The dynamics (24) is a special case of the dynamics (10) without influences from news agencies, i.e, $\hat{w}_{ik}(x_i(t)) = 0$ for $\forall i \in \mathcal{V}, \forall k \in \mathcal{K}$, $\forall t \in \mathbb{N}_0$. By Theorem 1, the dynamics (10) converges to an unique equilibrium point $x^{\text{ue}}$. Thus, at the steady state we have $x^{\text{ue}} = \hat{\alpha}s + Wx^{\text{ue}}$, from which (26) follows immediately. □

#### 4.1.2 In the Presence of News Agencies

*Corollary 2.* Consider the social dynamics (1), under Assumption 1. If (17) holds, the unique equilibrium point satisfies

$$x^{\text{ie}} = (\mathbf{1} - W)^{-1}\left(\alpha(x^{\text{ie}})s + \widetilde{W}(x^{\text{ie}})y\right), \quad (27)$$

where $W$, $\alpha(x^{\text{ie}})$, $s$, $\widetilde{W}(x^{\text{ie}})$ and $y$ are given by (15), (14), (11), (16) and (13), respectively.

*Proof:* It follows from (10) that at steady-state:

$$x^{\text{ie}} = \alpha(x^{\text{ie}})s + Wx^{\text{ie}} + \widetilde{W}(x^{\text{ie}})y. \quad (28)$$

The definition of $W$ in (15) with the condition (17) and (18) show that $W$ is a sub-stochastic matrix unless no individual in social network (1) is influenced by his innate opinion or any news agency. It is well-known that $\mathbf{1} - W$ is invertible if $W$ is a sub-stochastic matrix, thus (27) is obtained immediately from (28). □

Expanding $\alpha(x^{\text{ie}})s + \widetilde{W}(x^{\text{ie}})y$ by considering the definitions in (4), (14) and (16) yields

$$[\alpha(x^{\text{ie}})s + \widetilde{W}(x^{\text{ie}})y]_i$$
$$= \alpha_i(x_i^{\text{ie}})s_i + \sum_{k \in \mathcal{K}} \hat{w}_{ik}(x_i^{\text{ie}}, y_k)y_k$$
$$= (1 - \sum_{j \in \mathcal{V}} w_{ij} - \sum_{k \in \mathcal{K}} \hat{w}_{ik}(x_i^{\text{ie}}, y_k))s_i + \sum_{k \in \mathcal{K}} \hat{w}_{ik}(x_i^{\text{ie}}, y_k)y_k$$
$$= (1 - \sum_{j \in \mathcal{V}} w_{ij})s_i + \sum_{k \in \mathcal{K}} \hat{w}_{ik}(x_i^{\text{ie}}, y_k)(y_k - s_i). \quad (29)$$

From (25) and (11), we have

$$[\hat{\alpha} s]_i = (1 - \sum_{j \in \mathcal{V}} w_{ij}) s_i. \tag{30}$$

It follows from (29) and (30) that

$$[\alpha(x^{\text{ie}}) s + \widetilde{W}(x^{\text{ie}}) y]_i = [\hat{\alpha} s]_i + \sum_{k \in \mathcal{K}} \hat{w}_{ik}(x_i^{\text{ie}}, y_k)(y_k - s_i)$$

which is equivalent to

$$\alpha(x^{\text{ie}}) s + \widetilde{W}(x^{\text{ie}}) y = \hat{\alpha} s + \widetilde{W}(x^{\text{ie}}) y - \widehat{S}\widetilde{W}(x^{\text{ie}}) \mathbf{1}_m, \tag{31}$$

where

$$\widehat{S} = \text{diag}\{s_1, \cdots, s_n\} \in \mathbb{R}^{n \times n}. \tag{32}$$

Note that (27) subtracting (26) results in

$$x^{\text{ie}} - x^{\text{ue}} = (\mathbf{1} - W)^{-1}\left(\alpha(x^{\text{ie}}) s + \widetilde{W}(x^{\text{ie}}) y - \hat{\alpha} s\right). \tag{33}$$

Therefore, (31) and (33) imply

$$x^{\text{ie}} - x^{\text{ue}} = (\mathbf{1} - W)^{-1}(\widetilde{W}(x^{\text{ie}}) y - \widehat{S}\widetilde{W}(x^{\text{ie}}) \mathbf{1}_m). \tag{34}$$

Given the uninfluential equilibrium point $x^{\text{ue}}$ in (26), the relation (34) is useful in the estimation and Computation of the influential equilibrium point $x^{\text{ie}}$ in the following subsection.

### 4.2 Equilibrium Computation

Without knowledge of the state-dependent weight functions $g_{ik}(|x_i(t) - y_k|)$ in (4), the relation (34) implies a bound on the estimation of the equivalent point $x^{\text{ie}}$, or equivalently the bound on the deviation of influential equilibrium point $x^{\text{ie}}$ from uninfluential equilibrium point $x^{\text{ue}}$ caused by the presence of news agencies.

#### 4.2.1 Nonlinear Weight Functions

As Remark 2 states that $x_i(t) \in [0,1]$ for $\forall t \in \mathbb{N}_0$ and $\forall i \in \mathcal{V}$, we conclude that each weight function $\hat{w}_{ik}(x_i(t))$ given in (4) is bounded, therefore

$$\eta_i^l \leq \hat{w}_{ik}(x_i(t)) \leq \eta_i^u, \forall t \in \mathbb{N}_0, \forall i \in \mathcal{V}, \forall k \in \mathcal{K}. \tag{35}$$

**Theorem 2.** Consider the social dynamics (1) under the convergence condition (17). For any state-dependent weight function that satisfies (5) in Assumption 1 and its upper bound $\eta_i^u$ and lower bound $\eta_i^l$ in (35), the unique equilibrium point (11) satisfies

$$\Lambda^l \leq x^{\text{ie}} \leq \Lambda^u, \tag{36}$$

where

$$\Lambda^l = \max\left\{(\mathbf{1} - W)^{-1}(\Delta^l + \hat{\alpha} s), \mathbf{0}_n\right\}, \tag{37}$$

$$\Lambda^u = \min\left\{(\mathbf{1} - W)^{-1}(\Delta^u + \hat{\alpha} s), \mathbf{1}_n\right\}, \tag{38}$$

$$\Delta^l = \breve{\eta} B y - \widehat{S}\hat{\eta}\hat{n}, \tag{39}$$

$$\Delta^u = \hat{\eta} B y - \widehat{S}\breve{\eta}\hat{n}, \tag{40}$$

$$\breve{\eta} = \text{diag}\{\eta_1^l, \cdots, \eta_n^l\} \in \mathbb{R}^{n \times n}, \tag{41}$$

$$\hat{\eta} = \text{diag}\{\eta_1^u, \cdots, \eta_n^u\} \in \mathbb{R}^{n \times n}, \tag{42}$$

$$B = \begin{bmatrix} b_{11} & \cdots & b_{1m} \\ \vdots & \vdots & \vdots \\ b_{n1} & \cdots & b_{nm} \end{bmatrix} \in \mathbb{R}^{n \times m}, \tag{43}$$

$$\hat{n} = [\Gamma_1, \cdots, \Gamma_n]^\top \in \mathbb{R}^n, \tag{44}$$

with $\Gamma_i$, $\widehat{S}$, $\hat{\alpha}$, $s$ and $W$ given in (18b), (32), (25), (11) and (15), respectively.

*Proof:* Note that $\hat{w}_{ik}(x_i^{\text{ie}}) \geq 0$, from (35) and (4) one has $\eta_i^l \sum_{k \in \mathcal{K}} b_{ik} y_k \leq \sum_{k \in \mathcal{K}} \hat{w}_{ik}(x_i^{\text{ie}}) y_k \leq \eta_i^u \sum_{k \in \mathcal{K}} b_{ik} y_k$ for $\forall i \in \mathcal{V}$, which is equivalent to

$$\breve{\eta} B y \leq \widetilde{W}(x^{\text{ie}}) y \leq \hat{\eta} B y, \tag{45}$$

where $\breve{\eta}, \hat{\eta}, B, \widetilde{W}(x^{\text{ie}})$ and $y$ are given by (41), (42), (43), (16) and (13), respectively.

Note the definition of $\Gamma_i$ and $\hat{w}_{ik}(x_i^{\text{ie}})$ in (18b) and (4), respectively. From (35) one has $-s_i \eta_i^u \Gamma_i = -s_i \eta_i^u \sum_{k \in \mathcal{K}} b_{ik} \leq -s_i \sum_{k \in \mathcal{K}} w_{ik}(x_i^{\text{ie}}) \leq -s_i \eta_i^l \sum_{k \in \mathcal{K}} b_{ik} = -s_i \eta_i^l \Gamma_i$ for $\forall i \in \mathcal{V}$, which is equivalent to

$$-\widehat{S}\hat{\eta}\hat{n} \leq -\widehat{S}\widetilde{W}(x^{\text{ie}}) \mathbf{1}_m \leq -\widehat{S}\breve{\eta}\hat{n}, \tag{46}$$

where $\widehat{S}, \breve{\eta}, \hat{\eta}, \widetilde{W}(x^{\text{ie}})$ and $\hat{n}$ are given by (32), (41), (42), (16) and (44), respectively. Combining (45) with (46) yields

$$\Delta^l \leq \widetilde{W}(x^{\text{ie}}) y - \widehat{S}\widetilde{W}(x^{\text{ie}}) \mathbf{1}_m \leq \Delta^u, \tag{47}$$

where $\Delta^l$ and $\Delta^u$ are given by (39) and (40), respectively. Then from (34) and (47) one has

$$(\mathbf{1} - W)^{-1}\Delta^l \leq x^{\text{ie}} - x^{\text{ue}} \leq (\mathbf{1} - W)^{-1}\Delta^u. \tag{48}$$

Substituting the right-hand side of (26) into $x^{\text{ue}}$ in (48) results in

$$(\mathbf{1} - W)^{-1}(\Delta^l + \hat{\alpha} s) \leq x^{\text{ie}} \leq (\mathbf{1} - W)^{-1}(\Delta^u + \hat{\alpha} s). \tag{49}$$

It is known from Remark 2 that $x_i(t) \in [0,1]$ for $\forall t \in \mathbb{N}_0$ and $\forall i \in \mathcal{V}$, which implies that

$$\mathbf{0}_n \leq x^{\text{ie}} \leq \mathbf{1}_n. \tag{50}$$

Therefore, (36) follows from (49) and (50) immediately. □

**Remark 5.** With the only knowledge of bounds on weight functions (35), (36) can be viewed as the estimation of equilibrium point. If some entries, say $i$, of the right-hand side of (36) are equivalent to the corresponding entries $i$ of the left-hand side of (36), the estimation of equilibrium point of individuals $i$ is the precise $x_i^{\text{ie}}$, i.e, $[\Lambda^l]_i = x_i^{\text{ie}} = [\Lambda^u]_i$. The numerical example in Simulation section illustrates this point.

#### 4.2.2 Linear Weight Functions

Remark 2 states that the evolving individual opinions $x_i(t)$ at every time step $t \in \mathbb{N}_0$ belong to the small range set $[0,1]$. Hence, the nonlinear weight functions under Assumption 1 can be modeled as linear weight functions with small approximation errors.

The general linear weight functions considered in this subsection are described by:

$$\hat{w}_{ik}(x_i^{\text{ie}}) = \begin{cases} \beta_i - \gamma_i |x_i^{\text{ie}} - y_k|, & b_{ik} = 1 \\ 0, & b_{ik} = 0 \end{cases}$$

with $0 < \gamma_i < \beta_i < 1$, which can be rewritten equivalently as

$$\hat{w}_{ik}(x_i^{\text{ie}}) = (\beta_i - \gamma_i(x_i^{\text{ie}} - y_k) \text{sgn}(x_i^{\text{ie}} - y_k)) b_{ik} \tag{51a}$$

$$1 > \beta_i > \gamma_i > 0 \tag{51b}$$

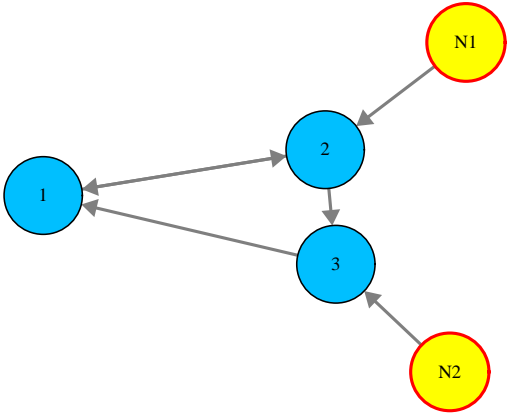

Figure 3. A social network: three individuals and two news agencies.

where $\mathrm{sgn}(\cdot)$ is defined as

$$\mathrm{sgn}(z) = \begin{cases} 1, & z > 0 \\ 0, & z = 0 \\ -1, & z < 0. \end{cases} \tag{52}$$

Under the linear weight functions (51), it is straightforward to verify from the definition of $\widetilde{W}(x^{\mathrm{ie}})$ in (16) with its entries defined in (51) that $\widetilde{W}(x^{\mathrm{ie}})$ can be rewritten equivalently as

$$\widetilde{W}(x^{\mathrm{ie}}) = \beta B + \gamma \widehat{W} \widehat{Y} - \gamma \widehat{X} \widehat{W}, \tag{53}$$

where $B$ is given in (43), and

$$\beta = \mathrm{diag}\{\beta_1, \cdots, \beta_n\} \in \mathbb{R}^{n \times n}, \tag{54}$$
$$\gamma = \mathrm{diag}\{\gamma_1, \cdots, \gamma_n\} \in \mathbb{R}^{n \times n}, \tag{55}$$
$$\widehat{X} = \mathrm{diag}\{x_1^{\mathrm{ie}}, \cdots, x_n^{\mathrm{ie}}\} \in \mathbb{R}^{n \times n}, \tag{56}$$
$$\widehat{Y} = \mathrm{diag}\{y_1, \cdots, y_m\} \in \mathbb{R}^{m \times m}, \tag{57}$$
$$\widehat{W} = \begin{bmatrix} \widehat{w}_{11} & \cdots & \widehat{w}_{1m} \\ \vdots & \vdots & \vdots \\ \widehat{w}_{n1} & \cdots & \widehat{w}_{nm} \end{bmatrix} \in \mathbb{R}^{n \times m}, \tag{58}$$
$$\widehat{w}_{ik} = b_{ik} \mathrm{sgn}(x_i^{\mathrm{ie}} - y_k), i \in \mathcal{V}, k \in \mathcal{K}. \tag{59}$$

Noting the defined matrix $\widehat{W}$ in (58) with its entries given in (59), we conclude that finite number of individuals and news agencies implies the finite number of possible $\widehat{W}$. To understand the meaning of finite number of possible $\widehat{W}$, consider the network of Figure 3 with three individuals and two news agencies. We observe from (59) with (52) and Figure 3 that

1) each $\widehat{w}_{ik}$ has three possible values: $\widehat{w}_{ik} = b_{ik}$, $\widehat{w}_{ik} = 0$ and $\widehat{w}_{ik} = -b_{ik}$;
2) news agencies $N_1$ and $N_2$ have two followers: individual 2 and individual 3.

Based on the above observations, we know that the social network in Figure 3 has $3^{1+1} = 9$ possible $\widehat{W}$:

$$\widehat{W}^1 = \begin{bmatrix} 0 & 0 \\ 1 & 0 \\ 0 & 1 \end{bmatrix}, \widehat{W}^2 = \begin{bmatrix} 0 & 0 \\ 1 & 0 \\ 0 & 0 \end{bmatrix}, \widehat{W}^3 = \begin{bmatrix} 0 & 0 \\ 1 & 0 \\ 0 & -1 \end{bmatrix},$$

$$\widehat{W}^4 = \begin{bmatrix} 0 & 0 \\ -1 & 0 \\ 0 & 1 \end{bmatrix}, \widehat{W}^5 = \begin{bmatrix} 0 & 0 \\ -1 & 0 \\ 0 & 0 \end{bmatrix}, \widehat{W}^6 = \begin{bmatrix} 0 & 0 \\ -1 & 0 \\ 0 & -1 \end{bmatrix},$$

$$\widehat{W}^7 = \begin{bmatrix} 0 & 0 \\ 0 & 0 \\ 0 & 1 \end{bmatrix}, \widehat{W}^8 = \begin{bmatrix} 0 & 0 \\ 0 & 0 \\ 0 & 0 \end{bmatrix}, \widehat{W}^9 = \begin{bmatrix} 0 & 0 \\ 0 & 0 \\ 0 & -1 \end{bmatrix}.$$

Generalizing the above example, the total number of possible $\widehat{W}$ is $3^{\sum_{k \in \mathcal{K}} \Gamma_k^{\mathrm{f}}}$ where $\Gamma_k^{\mathrm{f}}$ is the number of followers of new agency $k$, i.e., $\Gamma_k^{\mathrm{f}} = \sum_{i \in \mathcal{V}} b_{ik}$. It is a huge number if the social network has a lot of followers of news agencies. Fortunately, the estimation of equilibrium point in (36) can reduce the number of possible $\widehat{W}$ significantly. Using the estimation (36), from the definition of $\widehat{W}$ in (58) with (59) we define an ordered set of possible $\widehat{W}$ as follows.

$$\mathcal{W} = \left\{ \widehat{W}^1, \cdots, \widehat{W}^d, \cdots, \widehat{W}^p \right\}, \tag{60}$$

where $p$ is the total number of the possible $\widehat{W}$; the entries of each $\widehat{W}^d$ are defined by

$$\widehat{w}_{ik}^d = \begin{cases} -b_{ik}, & [\Delta^{\mathrm{u}}]_i < y_k \\ b_{ik}, & [\Delta^{\mathrm{l}}]_i > y_k \\ -b_{ik} \text{ or } 0, & [\Delta^{\mathrm{u}}]_i = y_k \\ b_{ik} \text{ or } 0, & [\Delta^{\mathrm{l}}]_i = y_k \\ -b_{ik} \text{ or } b_{ik} \text{ or } 0, & \text{otherwise.} \end{cases} \tag{61}$$

where $d = 1, 2, \cdots, p$, $\Lambda^{\mathrm{l}}$ and $\Lambda^{\mathrm{u}}$ are given by (37) and (38), respectively.

*Corollary 3.* For the linear state-dependent weight functions (51), the unique equilibrium point (27) can be solved by Algorithm 1.

*Proof:* From $\widehat{X}$ in (56) and $\widehat{S}$ in (32) one has $\widehat{X}\widehat{S} = \widehat{S}\widehat{X}$. Then it follows from (53) that

$$\widetilde{W}(x^{\mathrm{ie}})y - \widehat{S}\widetilde{W}(x^{\mathrm{ie}})\mathbf{1}_m \tag{66}$$
$$= (\beta B + \gamma \widehat{W}\widehat{Y} - \gamma \widehat{X}\widehat{W})y - \widehat{S}(\beta B + \gamma \widehat{W}\widehat{Y} - \gamma \widehat{X}\widehat{W})\mathbf{1}_m$$
$$= \widehat{X}(\gamma \widehat{S}\widehat{W}\mathbf{1}_m - \gamma \widehat{W}y) + (\beta B + \gamma \widehat{W}\widehat{Y})y - \widehat{S}(\beta B + \gamma \widehat{W}\widehat{Y})\mathbf{1}_m.$$

Let $\widehat{W}$ be the $d^{\mathrm{th}}$ element in the set $\mathcal{W}$ given by (60), i.e., $\mathcal{W}(d) = \widehat{W}$. Combining (34) with (66) yields

$$x^{\mathrm{ie}} - x^{\mathrm{ue}} = (\mathbf{1} - W)^{-1}(\widehat{X}G^a(d) + G(d)), \tag{67}$$

where $G^a(d)$ and $G(d)$ are respectively given by (64) and (65).

It follows from $\Theta(d)$ in (63) with (64), $\widehat{X}$ in (56) and $G^a(d)$ in (64) that $\widehat{X}G^a(d) = \Theta(d)x^{\mathrm{ie}}$. Then noting $x^{\mathrm{ue}}$ given in (26), from (67) one has $x^{\mathrm{ie}} - (\mathbf{1} - W)^{-1}\Theta(d)x^{\mathrm{ie}} = x^{\mathrm{ie}} - (\mathbf{1} - W)^{-1}\widehat{X}G^a(d) = x^{\mathrm{ue}} +$

**Algorithm 1:** Computation of Equilibrium Point

**Input**: Set $\mathcal{W}$ defined in (60) with the entries $\widehat{w}_{ih}^d$ of each element satisfying (61), initial index $d=1$, the number of elements of the set $\mathcal{W}$, i.e, $p$.

1 **while** $d \leq p$ **do**
2     Calculate:
$$x^{\text{ie}} = (\mathbf{1} - W - \Theta(d))^{-1}(G(d) + \hat{\alpha}s) \quad (62)$$
    where
$$\Theta(d) = \text{diag}\left\{[G^a(d)]_1, \cdots, [G^a(d)]_n\right\}, \quad (63)$$
$$G^a(d) = \gamma \widehat{S} \mathcal{W}(d) \mathbf{1}_m - \gamma \mathcal{W}(d) y, \quad (64)$$
$$G(d) \quad (65)$$
$$= (\beta B + \gamma \mathcal{W}(d)\widehat{Y})y - \widehat{S}(\beta B + \gamma \mathcal{W}(d)\widehat{Y})\mathbf{1}_m,$$
    with $\widehat{S}$, $B$ and $\widehat{Y}$ being given by (32), (43) and (57), respectively;
3     **if** $b_{ik}\text{sgn}\left(x_i^{\text{ie}} - y_k\right) = \widehat{w}_{ik}^d$, for $\forall i \in \mathcal{V}, \forall k \in \mathcal{K}$ **then**
4        Output equilibrium point: $x^{\text{ie}} \leftarrow x^{\text{ie}}$;
5        Break;
6     **else**
7        Update index: $d \leftarrow d+1$.
8     **end**
9 **end**

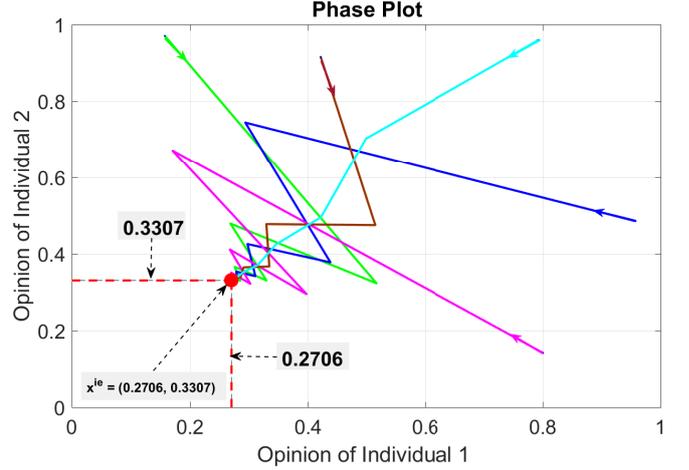

Figure 4. Computation of equilibrium point by Algorithm 1 and phase plot of opinions under five random initial opinions.

$(\mathbf{1} - W)^{-1}G(d) = (\mathbf{1} - W)^{-1}(G(d) + \hat{\alpha}s)$, which is equivalent to $(\mathbf{1} - (\mathbf{1} - W)^{-1}\Theta(d))x^{\text{ie}} = (\mathbf{1} - W)^{-1}(G(d) + \hat{\alpha}s)$. Therefore, we have

$$\begin{aligned} x^{\text{ie}} &= (\mathbf{1} - (\mathbf{1} - W)^{-1}\Theta(d))^{-1}(\mathbf{1} - W)^{-1}(G(d) + \hat{\alpha}s) \\ &= ((\mathbf{1} - W)(\mathbf{1} - (\mathbf{1} - W)^{-1}\Theta(d)))^{-1}(G(d) + \hat{\alpha}s) \\ &= (\mathbf{1} - W - \Theta(d))^{-1}(G(d) + \hat{\alpha}s). \end{aligned} \quad (68)$$

Note the loop stopping condition, i.e., Line 3 in Algorithm 1, is from the definition of entries of $\widehat{W}$ in (61). So (68) is, in fact, the Computation (62) in Algorithm 1. Theorem 1 states the equilibrium point $x^{\text{ie}}$ is unique. Computation (62) together with (63), (64) and (65) implies that once $\mathcal{W}(d) = \widehat{W}$ is searched, i.e., the condition of Line 3 in Algorithm 1 is satisfied, the equilibrium point $x^{\text{ie}}$ is solved, which completes the proof. □

*Example 4.* Consider the social network with two individuals and one news agency. Its communication structure is given in Figure 1 and its dynamics is given by (23). Set the individuals' innate opinions as $[s_1, s_2] = [0.1, 0.7]$. The opinion of the news agency $y_1 = 0.4$. The equilibrium point of the dynamics (23) calculated by Algorithm 1 is $x^{\text{ie}} = [0.2706, 0.3307]^\top$. The calculated equilibrium point and the phase plot of the dynamics (23) under five random initial opinions are shown in Figure 4, which shows the equilibrium point calculated by Algorithm 1 is correct.

We next focus on one particular special case for which we can analytically compute the equilibrium point. This is the setting where there is only once news agency and its belief is extreme. We define extreme opinion more formally in the following.

*Definition 1 (Extreme Opinion).* The opinion $\tilde{y}$ passed along by the only one news agency is said to be extreme with respect to other individuals' if

$$\tilde{y} \geq \max_{i \in \mathcal{V}} \{s_i\}. \quad (69)$$

Since the unique equilibrium point (27) is independent of initial opinions, it is also the equilibrium point of dynamics: $\tilde{x}(t+1) = \alpha(\tilde{x}(t))s + W\tilde{x}(t) + \widetilde{W}(\tilde{x}(t))\tilde{y}$ with $\tilde{x}(0) = \mathbf{0}_n$. Under the extreme opinion condition (69), in the situation that there is only one news agency it is straightforward to verify that $\tilde{y} \geq \max_{i \in \mathcal{V}} \{\tilde{x}_i(k)\}$, $\forall k \in \mathbb{N}_0$, which implies that $\tilde{y} \geq \max_{i \in \mathcal{V}} \{x_i^{\text{ie}}\}$.

*Corollary 4.* For the linear state-dependent weight functions (51) under the extreme opinion condition (69). In the situation that there is only one news agency who passes along the extreme opinion $\tilde{y}$, the unique equilibrium point (27) is solved as

$$x^{\text{ie}} = (\mathbf{1} - W - \Phi)^{-1}(H + \hat{\alpha}s), \quad (70)$$

where $\hat{\alpha}$ is given by (25), and

$$H = (\beta \breve{B} + \gamma \breve{B}\tilde{y})\tilde{y} - \widehat{S}(\beta \breve{B} + \gamma \breve{B}\tilde{y}), \quad (71)$$
$$\Phi = \text{diag}\left\{\gamma(\widehat{S} - \tilde{y}\mathbf{1})\breve{B}\right\}, \quad (72)$$
$$\breve{B} = [b_{11}, \cdots, b_{n1}]^\top, \quad (73)$$

with $\widehat{S}$, $\beta$ and $\gamma$ being given by (32), (54) and (55), respectively.

*Proof:* Note that under the extreme opinion condition (69) and the linear weight functions (51), from (16) and (51) the matrix $\widetilde{W}(x^{\text{ie}})$ can be rewritten equivalently as $\widetilde{W}(x^{\text{ie}}) = \beta\breve{B} + \gamma\breve{B}\tilde{y} - \gamma \widehat{X}\breve{B}$, where $\breve{B}$, $\widehat{X}$, $\beta$ and $\gamma$ are given in (73), (56), (54) and (55), respectively. Since the extreme opinion condition under linear weight functions is just a special case of general linear weight functions studied in the previous subsection, using the same analysis method to derive (62) in the proof of Corollary 3, one has (70). Hence

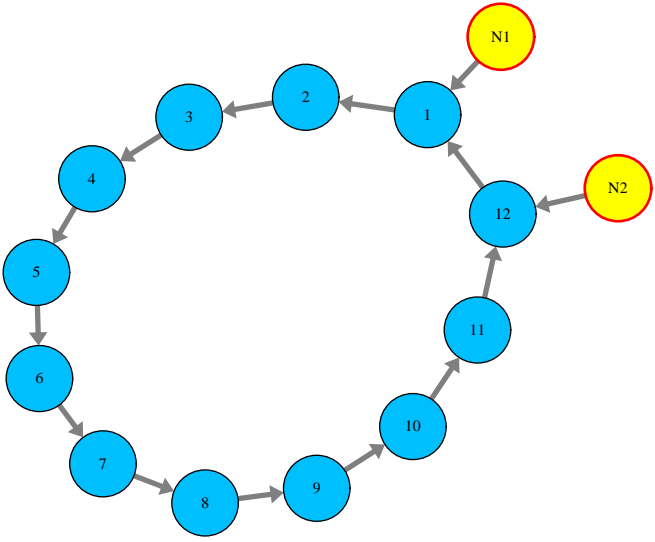

Figure 5. Ring topology with twelve individuals and two news agencies denoted by N1 and N2.

the remaining steps of proof follow from the those of the proof of Corollary 3 verbatim. □

*Remark 6.* It is straightforward to obtain from (70) and (26) that

$$\begin{aligned}x^{\text{ie}} - x^{\text{ue}} &= (\mathbf{1} - W - \Phi)^{-1}(H + \hat{\alpha}s) - (\mathbf{1} - W)^{-1}\hat{\alpha}s \\ &= \sum_{t=0}^{\infty}(W + \Phi)^t(H + \hat{\alpha}s) - \sum_{t=0}^{\infty}W^t\hat{\alpha}s \\ &\geq \sum_{t=0}^{\infty}W^t(H + \hat{\alpha}s) - \sum_{t=0}^{\infty}W^t\hat{\alpha}s \\ &= \sum_{t=0}^{\infty}W^tH = (\mathbf{1} - W)^{-1}H,\end{aligned}$$

which implies if the weight matrix $W$ given in (15) that describe interaction among individuals is a irreducible matrix, the news agency that forwards extreme opinion can drive the opinion of every individual away from the uninfluential equilibrium point.

## 5 SIMULATION

In this section, we first provide one numerical example to study the theoretical estimation of opinion evolution in the case of nonlinear weight functions. Then in the application to real social network, we investigate on the effects of the distribution of news agency's opinions and the distance between polar opinions of news agencies on opinion evolution.

### 5.1 Numerical Example

Consider the ring graph in Figure 5 that has twelve individuals and two news agencies. In the simulation setting: the innate opinions are randomly generated as $s = [0.9572, 0.4854, 0.8003, 0.1419, 0.4218, 0.9157, 0.7922, 0.9595, 0.6557, 0.0357, 0.8491, 0.9340]^\top$; the opinions of news agencies N1 and N2 are $y = [0.1, 0.8]^\top$; the weight matrix that describes friendship network in Figure 5 are set as $w_{12} = w_{23} = w_{34} = w_{45} = \frac{1}{2}$, $w_{56} = w_{67} = w_{78} = w_{89} = \frac{1}{3}$, $w_{9(10)} = w_{(10)(11)} = w_{(11)(12)} = w_{(12)1} = \frac{1}{4}$ and $w_{ij} = 0$ for other $i, j \in \mathcal{V}$; choose the state-dependent weight functions of individuals 1 and 2 the are influenced by new agencies N1 and N2 as

$$\hat{w}_{1(\text{N1})}(x_1(t)) = 0.4\ln(2 - |y_{\text{N1}} - x_1(t)|), \quad (74)$$
$$\hat{w}_{(12)(\text{N2})}(x_{12}(t)) = 0.4(1 - \sin|y_{\text{N2}} - x_{12}(t)|), \quad (75)$$

so the state-dependent weight $\breve{W}(x(t))$ defined in (16) under this setting is

$$\widetilde{W}(x(t)) = \begin{bmatrix} [\hat{w}_{1(\text{N1})}(x_1(t)), 0] \\ \mathbf{0}_{10\times 2} \\ [0, \hat{w}_{(12)(\text{N2})}(x_{12}(t))] \end{bmatrix}.$$

It follows from (7) and (9) that the nonlinear weight function (74) and (75) satisfy condition (5) in Assumption 1 with $\mu_1 = \mu_{12} = 0.4$; then it verifies that condition (17) in Theorem 1 is satisfied, thus the social network converges to a unique equilibrium point. Due to the nonlinearity of weight functions, it is difficult to calculate the equilibrium point manually. Considering (74) and (75) we have $0 \leq \hat{w}_{1(\text{N1})}(x_1(t)) \leq 0.4\ln 2$ and $0.4(1 - \sin(1)) \leq \hat{w}_{12(\text{N2})}(x_{12}(t)) \leq 0.4$. Then by Theorem 2, the upper bound and the lower bound of the estimation of equilibrium point in (36) are obtained as

$$\begin{aligned}\Lambda^{\text{l}} = [&0.4793, 0.5322, 0.5791, 0.3579, 0.5738, 0.8778, \quad (76)\\ &0.8020, 0.8217, 0.5460, 0.2171, 0.7612, 0.4975]^\top,\\ \Lambda^{\text{u}} = [&0.7725, 0.5323, 0.5791, 0.3579, 0.5739, 0.8782, \quad (77)\\ &0.8032, 0.8251, 0.5563, 0.2581, 0.9254, 1.0000]^\top.\end{aligned}$$

The trajectories of opinion evolution of six individuals are shown in Figure 6. The converged unique equilibrium point together with the lower bound and upper bound of the estimation of equilibrium point obtained in (76) and (77) are shown in Figure 7. We observe from Figure 6 and Figure 7 that

- compared with the individuals who are close to news agencies, the estimation errors of opinions of individuals that are far from news agencies are smaller;
- the estimations of the individuals' opinions at steady state approximate to their actual equilibrium points if the individuals are far way from the news agencies.

### 5.2 Real Network–Krackhardt's Advice Network

Based on the dynamics of cyber-social networks (1) proposed in this paper, we study the impact of the distribution of news agency's opinions and the distance between polarized opinions of news agencies on opinion evolution in the context of the well-known Krackhardt's advice network [20]. The communication topology of the aforementioned network is shown in Figure 8 with N1 indicting news agency.

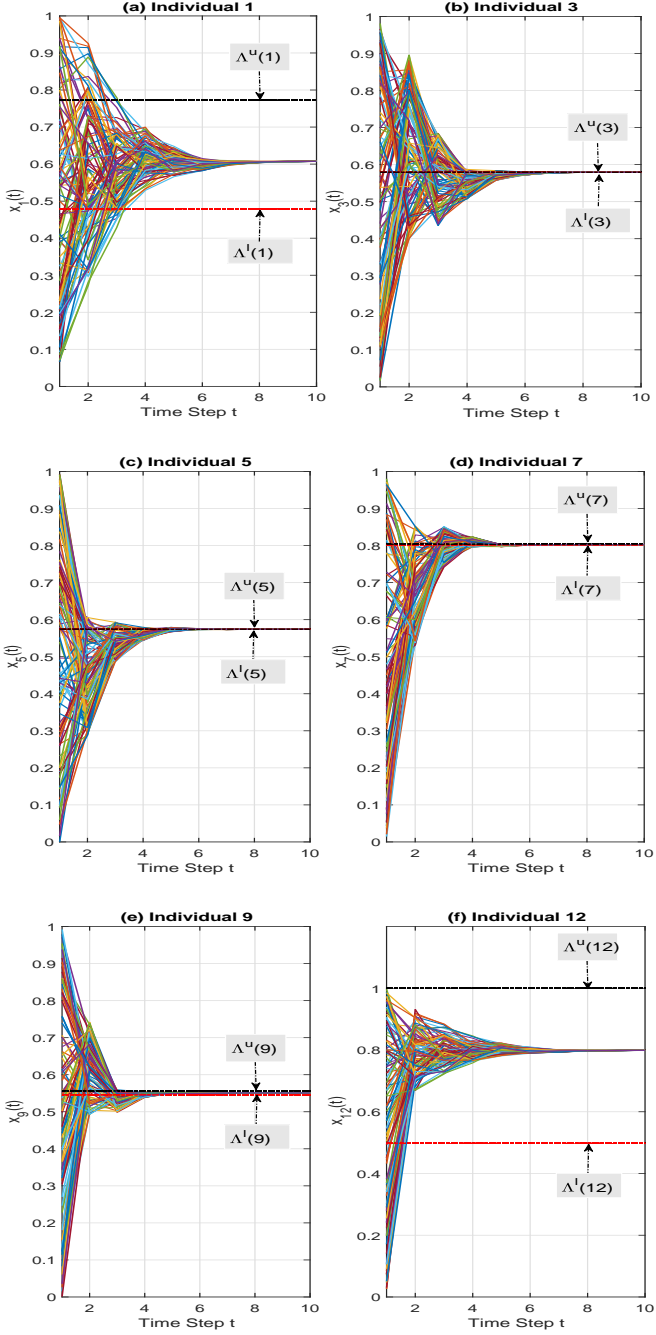

Figure 6. Trajectories of evolution of some individuals' opinions under 100 random initial opinions.

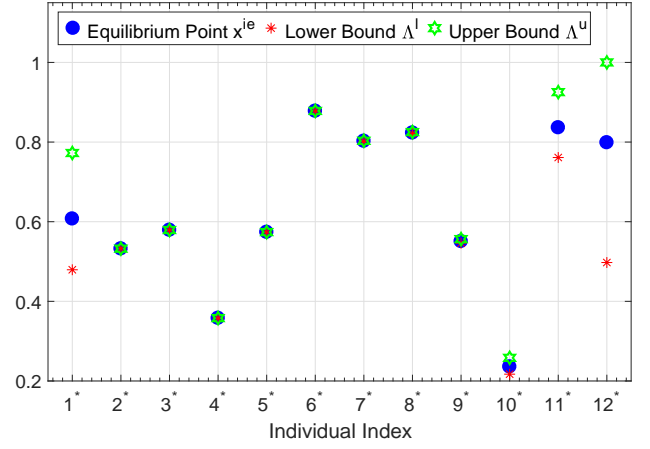

Figure 7. Equilibrium point and its estimation bounds calculated by Theorem 2.

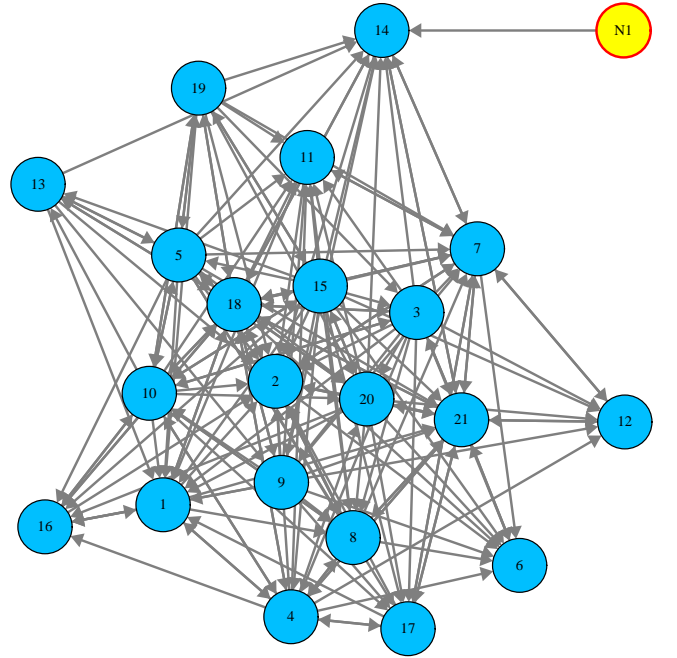

Figure 8. Krackhardt's advice network in the presence of new agency N1.

Using the source data available in [22], the outdegree distribution of the 21 individuals in Figure 8 are obtained as

$$\mathcal{N}^{\text{out}} = [\underbrace{6}_{1}, \underbrace{3}_{2}, \underbrace{15}_{3}, \underbrace{12}_{4}, \underbrace{15}_{5}, \underbrace{1}_{6}, \underbrace{8}_{7}, \underbrace{8}_{8},$$
$$\underbrace{13}_{9}, \underbrace{14}_{10}, \underbrace{3}_{11}, \underbrace{2}_{12}, \underbrace{6}_{13}, \underbrace{4}_{14}, \underbrace{20}_{15}, \underbrace{4}_{16},$$
$$\underbrace{5}_{17}, \underbrace{17}_{18}, \underbrace{11}_{19}, \underbrace{12}_{20}, \underbrace{11}_{21}]^{\top}, \quad (78)$$

For the weight matrix $W$ in (15), if individual $i$ asks for advice from his neighbor $j$, then $w_{ij} = \frac{1}{1+\Gamma_i^{\text{in}}}$ for all the individuals $j$ that influences individual $i$, where $\Gamma_i^{\text{in}} = \sum_{j \in \mathcal{V}} \text{sgn}(w_{ij})$. When an individual $i$ is influenced by the adversary news agency N1, his state-dependent weigh function is $\hat{w}_{i1}(x_i^{\text{ie}}) = \frac{1}{1+\Gamma_i^{\text{in}}} - \frac{1}{2(1+\Gamma_i^{\text{in}})}|x_i^{\text{ie}} - y_1|$.

In the following simulations, all of the 21 individuals' innate opinions follow the uniform distribution over $[0,1]$, i.e, $s_i \sim \mathrm{U}(0,1), \forall i \in \mathcal{V}$.

### 5.2.1 Distribution of News Agency's Opinions

To better see the effect of the distribution of news agency's opinions on opinion evolution, let the network has only one news agency, as denoted by N1 in Figure 8. Here, *the distribution of news agency* N1*'s opinion means which sole individual that news agency* N1 *intentionally influences*.

The effect of the presence of news agency N1 on the difference between influential equilibrium and uninfluential equilibrium point are measured by the sample deviation, which is defined as follows:

$$\sigma^2 = \frac{1}{n}\sum_{i=1}^{n} E_{\text{U}}((x_i^{\text{ue}} - x_i^{\text{ie}})^2). \qquad (79)$$

For the opinion of news agency N1, we consider two cases of uniform distributions:

$$\text{U}_1 : f(y) = \begin{cases} \frac{1}{0.8}, & y \in [0.1, 0.9] \\ 0, & \text{otherwise}, \end{cases} \qquad (80)$$

$$\text{U}_2 : f(y) = \begin{cases} \frac{1}{0.4}, & y \in [0, 0.2] \cup [0.8, 1] \\ 0, & \text{otherwise}, \end{cases} \qquad (81)$$

where $f(\cdot)$ is the probability density function.

Note that the means of the two uniform distributions in (80) and (81), and the distribution of innate opinion are as the same as 0.5. With 10,000 random samples, the standard sample deviations $\sigma$ under the two different uniform distributions are shown in Figure 9 (a), *where individual index $i^o$ denotes news agency* N1 *influences individual $i$ solely*. Figure 9 (a) together with Figure 9 (b) or (78) show that

- for news agency, influencing critical individual, i.e, the individual with highest outdegree, results in biggest sample deviation;
- the order of influences of individuals that follow news agency on sample deviation is preserved under different uniform distributions;
- bi-model uniform distribution ($\text{U}_2$ in (81)) results in bigger sample deviation than single-model uniform distribution ($\text{U}_1$ in (80)).

### 5.2.2 Distance Between Polar Opinions

The Krackhardt's advice network that is in the presence of two news agencies that pass along polar opinions is shown in Figure 10. As Figure 10 shows the news agencies N1 and N2 pass along their opinions to all of the 21 individuals. To study the effect of the distance between the two polar opinions of news agencies N1 and N2 on the opinion evolution, we use sample variance to measure the influences of different polar opinions, which is defined as follows:

$$\delta^2 = \frac{1}{n}\sum_{i=1}^{n} E_{\text{U}}((x_i^{\text{ie}} - \bar{x})^2), \qquad (82)$$

where $\bar{x}$ is the average of group opinions, i.e., $\bar{x} = \frac{1}{n}\sum_{i=1}^{n} x_i^{\text{ie}}$.

The two news agencies N1 and N2 consider six uniform distributions, which are given in Figure 11. Note that the means of polar opinions and the innate opinions are as the same as 0.5. With 10,000 random samples, the standard sample variances $\delta$ under six different uniform distributions are shown in Figure 12. Figures 12 and 11 show that the longer distance between the means of polar opinions results in the larger sample variance.

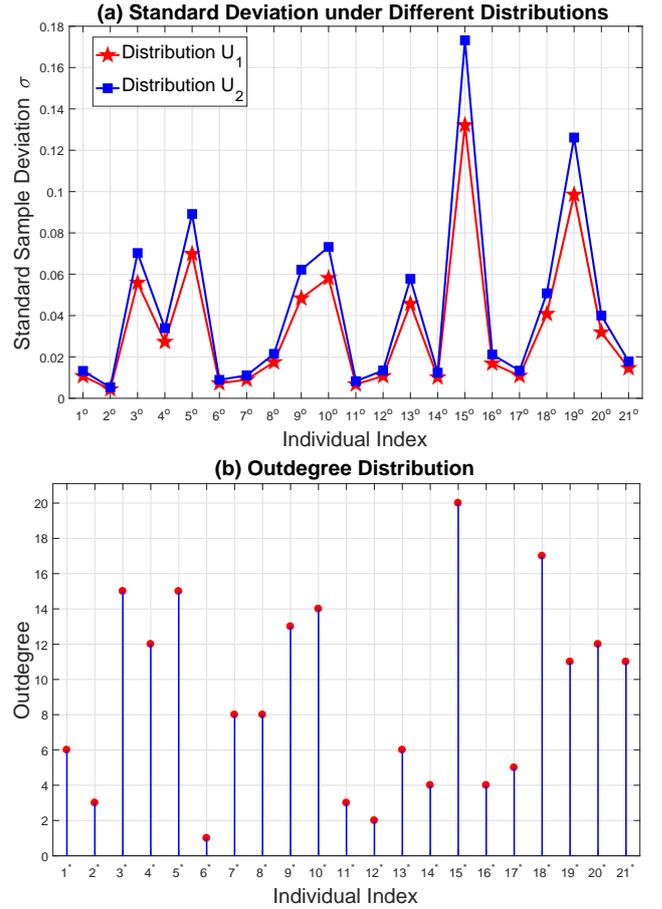

Figure 9. Standard sample deviation under different uniform distributions, and outdegree distribution of 21 individuals.

## 6 CONCLUSION

This paper studies the dynamics of information spread on cyber-social networks. The dynamics is adopted from the well-known DeGroot-Friedkin model, where the evolution of an individual's opinion at each time step is modeled as a convex combination of her innate opinion, her neighbors' opinions at the previous time step, and the opinions forwarded by the news agencies in the cyber layer which she follows. The weights are determined by a weight function that is dictated by the characteristics of the confirmation bias. Using the well-known Banach fixed-point (contraction) theorem, we characterize the conditions for convergence to a unique equilibrium (steady-state) point, which is independent of initial opinions. The steady-state points of the proposed social dynamics under both linear and nonlinear weight functions are studied. The estimation of equilibrium point is derived for nonlinear weight functions. An algorithm that precisely computes the equilibrium point for linear weight functions is provided. Theoretical results are verified by numerical examples.

Based on our model, we numerically analyze the impact of polarized news agencies. The simulation results show that

1) the order of influences of individuals that follow news agency on sample deviation is preserved under differ-

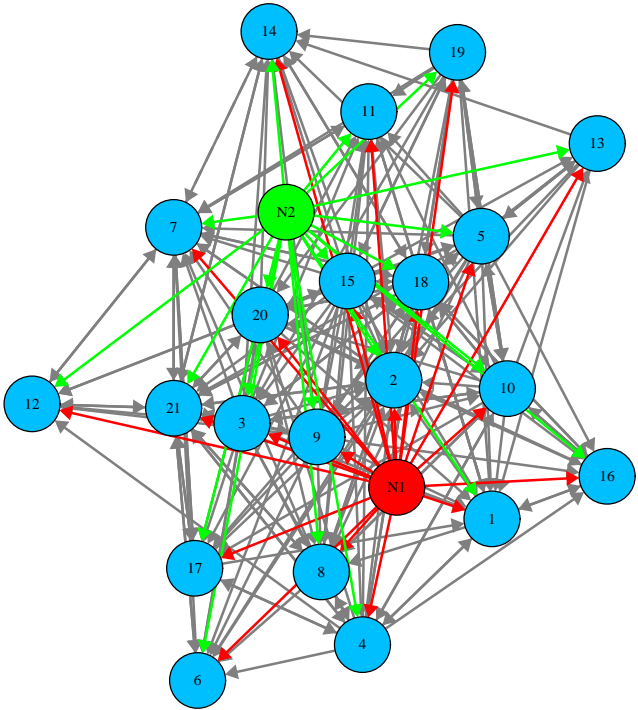

Figure 10. Krackhardt' advice network in the presence of two news agencies $N_1$ and $N_2$.

ent distributions;
2) bi-modal uniform distribution results in larger sample deviation than unimodal uniform distribution;
3) the larger distance between the means of polarized opinions results in the larger sample variance.

The obtained equilibrium points show the innate opinions play a critical role in the expressed opinions of individuals. Therefore, more transparent innate opinion implies less individual privacy and more precise prediction of public opinion. The equilibrium points together with simulations on Krackhardt's advice network provide the following game-theoretic insights:

- how individuals should strategically express his opinions to leak least information of innate opinions to adversary news agencies, while how adversary news agencies should strategically deploy observers to learn most information about the innate opinions,
- how news agencies should strategically express their opinions (polarized opinion distributions) to influence most individuals, while how individuals should strategically follow opinions of news agencies (weights of influence of news agencies) to accept the least extreme opinions.

Analyzing our model in the light of game theory, where different subset of news agencies are controlled by different players, contitutes a part of our future research.

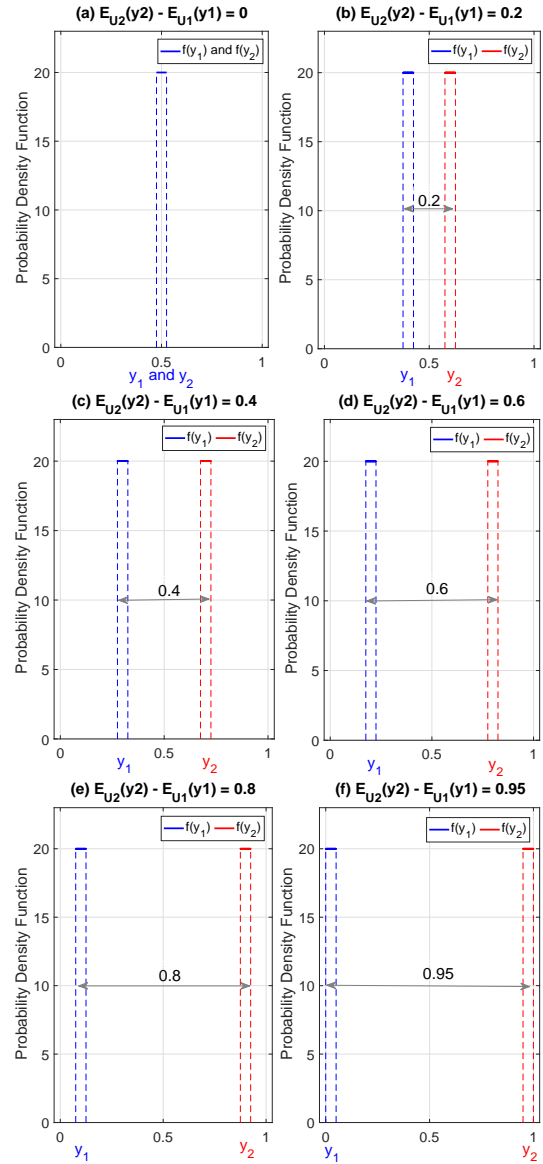

Figure 11. Uniform distributions of polar opinions: mean of each distribution is 0.5.

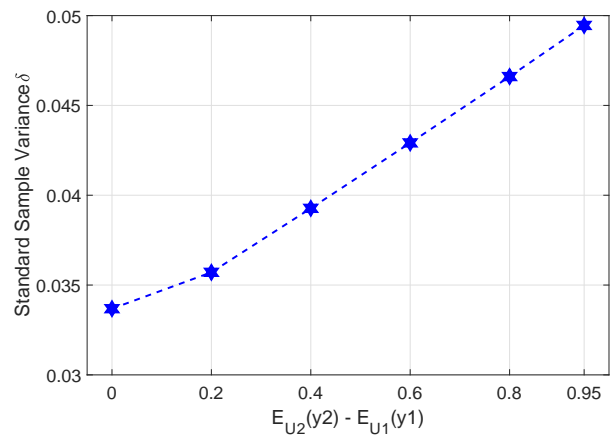

Figure 12. Standard sample variance under different uniform distributions.